\documentstyle[preprint,version2,aps]{revtex}
\newcommand{\vect}[1]{\mbox{\boldmath ${#1}$}}
\begin{document}
\draft
\rightline{cond-mat/9401053}
\begin{title}
Numerical Study of the $S=1$ Antiferromagnetic Spin Chain\\
with Bond Alternation\\
\end{title}
\author{Yusuke Kato and Akihiro Tanaka}
\begin{instit}
Department of Applied Physics, University of Tokyo, Bunkyo-ku,
Tokyo 113, Japan
\end{instit}
\newcommand{\bold}[1]{\mbox{\boldmath ${#1}$}}
\newcommand{\PRB}[3]{Phys. Rev. B {\bf {#1}} ({#3}) {#2}}
\newcommand{\PRL}[3]{Phys. Rev. Lett. {\bf {#1}} ({#3}) {#2}}
\newcommand{\PLS}[3]{Phys. Lett. {\bf {#1}} ({#3}) {#2}}
\newcommand{\PR}[3]{Phys. Rev. {\bf {#1}} ({#3}) {#2}}
\newcommand{\PC}[3]{Physica C {\bf {#1}} ({#3}) {#2}}
\newcommand{\JLTP}[3]{J. Low Temp. Phys. {\bf {#1}} ({#3}) {#2}}
\newcommand{\JAP}[3]{J. Appl. Phys. {\bf {#1}} ({#3}) {#2}}
\newcommand{\ZETF}[3]{Zh. Eksp. Teor. Fiz. {\bf {#1}} ({#3}) {#2}}
\newcommand{\ZP}[3]{Z. Phys. {\bf {#1}} ({#3}) {#2}}
\newcommand{\APL}[3]{Appl. Phys. Lett. {\bf {#1}} ({#3}) {#2}}
\newcommand{\JP}[3]{J. Phys. {\bf {#1}} ({#3}) {#2}}
\newcommand{\RMP}[3]{Rev. Mod. Phys. {\bf {#1}} ({#3}) {#2}}
\newcommand{\rot}{\mbox{rot}}
\newcommand{\red}{\rm{red}}
\newcommand{\fref}[1]{Fig. \ref{#1}}
\newcommand{\sign}{sgn}
\begin{abstract}
(RECEIVED \hspace{9cm})\\

 We study the $S=1$ quantum spin chain with bond alternation ${\cal H}=\sum _i
(1-(-1)^i\delta)\vect{S}_i\cdot \vect{S}_{i+1}$ by the density matrix
renormalization group method recently proposed by Steven R. White
(\PRL{69}{3844}{1993}).
 We find a massless point at $\delta _c =0.25 \pm 0.01$. We also find the edge
states in the region $\delta <\delta_c$ under the open boundary condition,
which disappear in the region $\delta >\delta _{c}$. At the massless point, the
spin wave velocity $v_s$ is $3.66 \pm 0.10$ and the central charge $c$ is
$1.0\pm 0.15$. Our results indicate that a continuous phase transition occurs
at the massless point $\delta =\delta_c $ accompanying breaking of the hidden
$Z_2\times Z_2$ symmetry.
\end{abstract}
The antiferromagnetic $S=1$ Heisenberg chain has been the subject of a large
amount of interest in the last decade since Haldane argued the difference of
the low energy physics between integer and half-integer spin
\cite{Haldane,Haldane12}. The key concept of the Haldane conjecture is
described in terms of the $O(3)$ nonlinear sigma model with the topological
term and he concluded that the topological term is ineffective and the spectrum
is massive for integer spin, while the topological term suppresses the
mass-generation and excitation spectrum is massless for half-integer spin. For
the $S=1/2$ Heisenberg model, it is established that the excitation spectrum is
gapless by the Bethe-Ansatz solution. The existence of excitation gap in the
case of $S=1$ was verified from the theoretical \cite{AKLT,AKLT2}, numerical
\cite{Botet,Botet2,Botet3,Nightingale,Sakai}, and experimental results
\cite{Renard,Renard2,Renard3,Renard4}.

Some years later, Haldane and Affleck
\cite{Haldane2,Affleck1,Affleck2,Affleck22,AffleckHaldane} considered the
antiferromagnetic (AF) spin chain with bond alternation $\delta$,
\begin{equation}
{\cal H}=\sum
_{i=1}^{N_s-1}\left(1-\left(-1\right)^i\delta\right)\vect{S}_i\cdot
\vect{S}_{i+1}.\label{eq:alter}
\end{equation}
When this Hamiltonian is mapped to the continuum model, {\it i.e.}, the O(3)
nonlinear sigma model, the topological angle $\theta$, which is the coefficient
of the Berry phase term in the action, is given by $\theta=2\pi S(1-\delta)$.
According to the argument by Haldane and Affleck, only in the case where
$\theta=\pi\times(\mbox{odd integer})$, is the system in the critical state
with massless excitations. Hence as the bond alternation $\delta$ changes from
$-1$ to $1$, the system experiences critical (massless) point $2S$ times.
Applied to the $S=1/2$ case, this general discussion is consistent with
Cross-Fisher's argument of the spin-Peierls transition \cite{CrossFisher}.
However the situation is nontrivial in the case of $S=1$. Without the bond
alternation $\delta$, the system is gapful, {\it i.e.}, possesses the Haldane
gap, and the above discussion predicts that the massless point appears at some
$\delta=\delta_c \ne 0$. However this argument contains several approximations
and assumptions, such as taking the large $S$ limit, and it is important to
study the model (\ref{eq:alter}) by another approach and to give an independent
check of the above scenario.

Varying-$\delta$ problem in the model $(1)$ can be attacked from the viewpoint
of the topological order proposed by den-Nijs and Rommelse \cite{dennijs} and
Tasaki \cite{Tasaki}. This ^^ ^^ hidden" order is measured by the string
correlation function
\begin{equation}
g\left(i,j\right)=\langle S_i^z \left(\prod ^{j-1}_{k=i+1}e^{i\pi
S_k^z}\right)S_j^z\rangle. \label{string}
\end{equation}
It is widely believed that this correlation function is long-ranged at the
Heisenberg point $\delta=0$ from numerical evidences
\cite{GirvinArovas,WhiteHuse}. On the other hand, an elementary calculation
shows that the string correlation function is exactly zero at the completely
dimerized points $\delta =\pm 1$. Hence we expect that phase transitions
corresponding to the hidden symmetry breaking should occur twice; once in the
interval between $\delta =-1$ and $\delta=0$ and once more between $\delta =0$
and $\delta =1$. However, these arguments cannot determine whether the phase
transitions are discontinuous or continuous. Thus the existence of the massless
points as predicted by Haldane and Affleck is still an open problem and direct
numerical study is needed.
In this paper we investigate the model (\ref{eq:alter}) with $S=1$ numerically
for the region $0 \leq \delta \leq 1$, using exact diagonalization and the
density matrix renormalization group (DMRG) method. This is sufficient since
reversing the sign of $\delta$ merely corresponds to interchanging the roles of
the two sublattices, and hence has no effects on the bulk properties. It does
however have an intrinsic boundary effect for open chains on which we will
remark later.

DMRG, which was recently proposed by White, is a new algorithm for real-space
renormalization group \cite{WhiteHuse,White,Sorensen,White2}. The key point
lays in the method of the truncation procedure of the Hilbert space.
In conventional real space renormalization group calculation, one keeps the
lowest-lying eigenstates of the block Hamiltonian in forming a new effective
Hamiltonian.
In the DMRG method, on the other hand, one keeps the most important eigenstates
of the block density matrix, obtained from diagonalizing the Hamiltonian of a
larger section of the lattice which includes the block. The above change brings
high accuracy and the new algorithm is already applied to the S=1
antiferromagnetic Heisenberg spin chain successfully
\cite{WhiteHuse,White,Sorensen,White2}. For details of the DMRG algorithm,
readers are referred to ref.\cite{White2}.

 Using the DMRG method, we calculated the excitation gap of systems of large
length up to about two hundred sites and extrapolated them to the thermodynamic
limit. The numbers of state kept per block in our calculations are 27, 36, 48,
and 81. As a benchmark test we calculated the ground state energy density of
$S=1/2$ Heisenberg model. Comparing with the exact Bethe-Ansatz solution, we
find that the relative error is about $5 \times 10^{-4}$. We also compared
those of DMRG with that of exact diagonalization in the case of $S=1$ and site
number $N_s=12$. The two results were found to agree well within a relative
error of $7 \times 10^{-6}$.

Figure 1 shows the excitation spectrum of $S=1$ spin chain versus dimer
strength. The data of diagonalization
(points plotted with triangles, squares, empty and filled circles and
empty and filled diamonds) shows the gap between the singlet ground state and
the first excited (triplet) state under periodic boundary condition.
The results of diagonalization show that the gap has a minimum around $\delta
\sim 0.25$, which has been already reported in ref. \cite{Guo}. The minimum
value of the gap becomes smaller with increasing system size. In the large
$N_s$ limit, there are two possibilities {\it i.e.} the minimum value of the
gap tends to zero or saturates at a finite value. Limited system sizes in the
diagonalization, however, disenables us to conclude which is realized in the
thermodynamic limit.

The data of DMRG calculation (represented by the double circles and crosses),
which are extrapolated values ($N_s\rightarrow +\infty$), show the gap between
the ground state and the first excited state, and that between the ground state
and the second excited state under the open boundary condition. The gap
behaviors are different between the region $\delta <0.25$ and $\delta >0.25$.
In the region $\delta <0.25$, the first excited state is triplet $(S_{{\rm
tot}}=1)$ and the second one is quintuplet $(S_{{\rm tot}}=2)$ while both
states are triplet in the region $\delta >0.25$. The first excited state in the
region $\delta <0.25$, approaches the singlet ground state exponentially with
increasing system size. These triplet excitations have been investigated at the
Heisenberg point ($\delta =0$) and are known to be the edge excitations near
the open boundary of the spin chain \cite{Kennedy,Hagiwara}. For the moment, we
focus upon the gap between ground state and the second excited state in the
region $\delta <0.25$, because the above ^^ ^^ edge excitations" seem to have
little effects upon the bulk properties of the spin chain. Later we will
discuss the edge states, especially their relevance to the hidden symmetry
breaking.

 The size dependence of the gap between the ground state and the second excited
state in the region $\delta < 0.25$ is fitted well with $1/N_s^2$. The gap
between the ground state and the first and second excited states in the region
$\delta > 0.25$ is also fitted well with $1/N_s^2$. At $\delta =0.25$, size
dependence of the gap behaves as $1/N_s$.
In contrast to the exact diagonalization method, DMRG gives a clear evidence
for the existence of a massless point at $\delta _c=0.25 \pm 0.01$.

We now turn to the critical properties of the massless point. It is plausible
to assume that at the massless point, the system has conformal invariance with
respect to the low energy properties. Actually, we observe that the size
dependence of the excitation gap is proportional to $1/N_s$, which is
consistent with the finite size scaling of the conformal field theory (CFT).
 CFT gives the expression for the size dependence of ground state energy
\cite{Blote,Blote2};
\begin{equation}
\frac{E_{g.s}(N_s)}{N_s}=\epsilon + \frac{f}{N_s}-\frac{\pi c
v_{s}}{24N_s^2}+{\cal O}(\frac{1}{N_s^3}),
\end{equation}
 under the open boundary condition. Here $E_{g.s}(N_s)$, $\epsilon$, $f$, and
$c$ are the ground state energy of finite size system, ground state energy
density of the infinite system, surface energy and the central charge.
 We estimate $v_s =3.66 \pm 0.10$ at $\delta=\delta_c$ from the excitation gap.
 Combining the value of $v_s$ with the results for the ground state energy, we
estimate $c = 1.0\pm 0.15$.

Affleck and Haldane predicted $\delta_c=0.5$, which deviates from our result
$\delta_c =0.25 \pm0.01$. We can attribute this discrepancy to the mapping of
the quantum spin chain onto the O$(3)$ sigma model, which is justified in the
large $S$ limit. However their prediction is correct with regards to the {\it
existence} of a massless point.

 The existence of edge states or four-fold degeneracy of the ground state in
the open chain was discussed theoretically
\cite{AKLT,AKLT2,Kennedy,KennedyTasaki} and was related to the long range
topological order and hidden $Z_2 \times Z_2$ symmetry breaking. So we conclude
that at the massless point a continuous phase transition with the hidden $Z_2
\times Z_2$ symmetry breaking occurs.
Our result ($c \sim 1$) is in accord with the general belief concerning the
$Z_2\times Z_2$ symmetry breaking \cite{Ashkin} that the massless phase
boundary which corresponds to a full $Z_2\times Z_2$ symmetry breaking belongs
to the same universality class as the Gaussian model.

We note in passing the consequence of our choice of the sign of $\delta$; had
we taken negative $\delta$, we would have found that the ground state
degeneracy is nine-fold in the region $-1 \leq \delta \leq \delta_c^- $ (due to
the formation of two localized $S=1$ objects at the open ends), while it is
still four-fold in the region $\delta_{c}^{-} \leq \delta <0$, where
$\delta_c^{-} \equiv -\delta_c \sim -0.25$.

In summary, we presented numerical results on the $S=1$ antiferromagnetic spin
chain with bond alternation and concluded that the continuous phase transition
occurs at $\delta_c =0.25\pm 0.01$ with the hidden $Z_2\times Z_2$ symmetry
breaking. The higher $S$ cases are left for future investigations.

We are grateful to Professor N. Nagaosa for introducing us to this problem and
for continuous encouragement as well as for critical reading of the manuscript.
This work is supported by a Grant-in-Aid for Scientific Research on Priority
Areas, ^^ ^^ Computational Physics as a New Frontier in Condensed Matter
Research" from the Ministry of Education, Science, and Culture of Japan.
Numerical calculations were performed on HITAC S-3800 at the computer center of
University of Tokyo. We also acknowledge Professor M. Kaburagi, Professor T.
Tonegawa, and Dr. T. Nishino for their diagonalization program KOBEPACK/1.

\figure{Excitation gap between the ground state and excited states versus the
dimer strength $\delta$. We show here two kinds of numerical results.
Points plotted with triangles, squares, circles (empty and filled) and
diamonds (empty and filled)
are the gap between the ground state and the first excited state obtained by
the exact diagonalization under the periodic boundary condition for system
sizes $N_s=4, 6, 8, 10, 12$, and $14$. As the system becomes larger, the
minimum of the gap shows up around $\delta \sim 0.25$. Double circles/crosses
represent the gap between the ground state and the first/second excited state
obtained by the DMRG method (the infinite lattice method) under the open
boundary condition. We find a massless point around $\delta_c=0.25 \pm 0.01$,
and also edge states (Kennedy triplet) in the region $\delta <\delta_c$.
}
\end{document}